\documentstyle[12pt,psfig,epsfig]{article}
\def\be{\begin{equation}}
\def\ee{\end{equation}}
\def\ba{\begin{eqnarray}}
\def\ea{\end{eqnarray}}
\def\ga{\mathrel{\raise.3ex\hbox{$>$\kern-.75em\lower1ex\hbox{$\sim$}}}}
\def\la{\mathrel{\raise.3ex\hbox{$<$\kern-.75em\lower1ex\hbox{$\sim$}}}}
\newcommand{\sect}[1]{\section{#1}\setcounter{equation}{0}}

\newcommand{\fr}[2]{\frac{#1}{#2}}

\begin{document}   

\begin{titlepage} 
\rightline{hep-ph/yymmnn}
\rightline{December 2004}  
\begin{center}

\vspace{0.5cm}

\large {\bf Ghosts and Tachyons in the Fifth Dimension}\\[3mm]
  
\vspace*{0.5cm}
\normalsize

{\bf Maxim Pospelov} \footnote{On leave of absence from:
{\it Dept. of Physics \& Astronomy, 
University of Victoria,
Victoria, BC, V8W 3P6, Canada  } }

\smallskip 
\medskip 

{\it Perimeter Institute, 31 Caroline Street North, 
Waterloo, ON,  N2J 2W9,
Canada
}

and 

{\it  Department of Physics,
 University of Guelph,
 Guelph, ON,  N1G 2W1, Canada
 } 
%$^{\,(c)}$
\smallskip 
\end{center} 
\vskip0.6in 
 
\centerline{\large\bf Abstract}
We present several solutions for the five dimensional 
gravity models in the presence of bulk ghosts and tachyons 
to argue that these "troublesome" fields can be a useful
model-building tool. 
The ghost-like signature of the kinetic term for a bulk scalar 
creates a minimum in the scale factor,
removing the necessity for a negative tension brane in models 
with the compactified fifth dimension. It is shown that the model with 
the positive tension branes and a ghost field in the bulk leads to 
the radion stabilization. The bulk scalar with the variable sign 
kinetic term can be used to model both positive and negative 
tension branes of a finite width  in  the compact dimension. 
Finally, we present several ghost and tachyon field configurations 
in the bulk that lead to the localization of gravity
in four dimensions, including one solution with the Gaussian profile for the metric,
$g_{\mu\nu}(y)=\eta_{\mu\nu}\exp\{-\alpha y^2\}$,  which leads to a stronger localization 
of gravity than the Randall-Sundrum model.

\vspace*{2mm} 
%\smallskip\newline

\end{titlepage}

\sect{Introduction}

Internal compactified space-like dimensions 
are an important ingredient of the string theory, 
that recently made an impressive comeback to the 
fields of theoretical particle physics and cosmology \cite{HW,reviewpp,ADD,RS,hierarchy,cosmo1}. 
Following the work of Randall and Sundrum (RS) \cite{RS}, an enormous amount of 
literature has been devoted to studies of five dimensional models with 
a nontrivial dependence of the scalar factor $a$ on internal coordinate 
$y$, $a = a (y)$. Various extensions of the RS idea included the study 
of the dynamics of the model in the presence of higher--dimensional gravitational
interaction terms \cite{Gauss-Bonnet}, 
bulk fields \cite{GW}, quantum effects \cite{casimir}, 
brane--induced kinetic terms \cite{DGP}, etc. The addition of such 
"extra features" may lead to the stabilization of an 
extra dimension \cite{GW,kkop}, can provide a glimpse of
hope for the solution to the cosmological constant problem \cite{Stanford,Shifman}, result in a 
vivid 5D--derived particle physics phenomenology (see {\em e.g.} \cite{pheno} and 
references therein), and so on. In this paper we advocate a point of 
view that even more exotic possibilities such as bulk ghosts and tachyons 
may be a useful "engineering" tool in the fifth dimension. 

The original RS model \cite{RS} employed the negative tension brane at the 
minimum of the scale factor $a(y)$. Such object, a well-localized 
amount of negative energy, looks counter-intuitive as it has no analogs 
in conventional physics (although it may find some support in string theory). 
Models of compact dimension that involve only positive 
tension branes have to come up with some alternatives to negative tension 
brane in order to ensure the turn-around of the scalar factor at some 
$y_{min}$, i.e. $a'(y_{min})=0$ and $ a''(y_{min})>0$. Conventional 
fields in the bulk are not able to produce 
such a solution at the classical level. Known static solutions 
with the $a \sim \cosh(\omega y)$ behavior 
necessarily involve the stress-energy tensor in the bulk that violates 
weak energy condition in one way or another. Such solutions include 
positive tension brane models in the presence of quantum effects such as Casimir force
\cite{casimir} or some solutions with higher-derivative gravitational terms \cite{Gauss-Bonnet}.
Cosh-like solution also arise in the time-dependent 
Randall-Sundrum model when the brane tension is
detuned from a flat case, resulting in 4D AdS geometry \cite{KR}.  

In this paper we generalize the results of \cite{kop1,kop2} 
that find the solutions to the scale factor and analyze the radion stability 
in the 5D models with conventional bulk scalar.
We consider models that admit scalar fields that has the "wrong" sign of the kinetic term. 
Such fields, termed ghost hereafter, can be a source of 
de-localized negative energy and assume the role of 
an effective  negative tension brane. Indeed, it is easy to show 
that the bulk ghost leads to a $\cosh$-like solution and that this solution 
is stable against the perturbation in the size of an extra dimension. 
This does not insure the full stability because of  
inherent instability of models with ghost fields that has to be dealt with separately. 
A more generic case of a field with the variable sign kinetic term can be used 
to construct positive energy -- negative energy kink--anti-kink configurations, 
that in the limit of small widths of such defects can be considered as positive 
tension--negative tension branes. 

We also discuss how various scalar field species ({\em i.e.} regular scalar fields, ghost and 
tachyons) can be used to localize gravity in a non-compact dimension. It is shown that several models 
lead to either $\exp\{-|y|\}$ or $\exp\{-y^2\}$ asympototics of the scale factor, 
rendering a finite integrals over the extra dimension and 4D gravity localization.
The Gaussian-like profile for an extra dimension can arise as a limiting case of certain 
5D theories with higher-derivative gravitational terms included in the action \cite{Bob},
and we would like to investigate if simpler realizations are also possible.
Some of the models found in this paper do not require any presence of branes and in certain cases result in 
the flat 4D solutions that can be found without an additional tuning of the parameters in the Lagrangian.

At this point we would like to recall that the idea to use ghosts in an extra-dimensional 
set-up has been suggested before. In particular, Ref. \cite{kkop} argued 
that the non-linear Lagrangian for a scalar field $\phi$
\begin{eqnarray}
{\cal L}_{\rm scalar} = -{\rm const} \times (\partial_M \phi \partial^M\phi  - c_1^2)^2 
\label{intro}
\end{eqnarray}
can be used for a radion stabilization in the 5D model. 
Here the term linear in $\partial_M \phi \partial^M\phi$
has a sign opposite to a conventional sign. Notice, that throughout this paper 
we employ $\small{-++++}$ signature for the metric.
The proposal of Ref. \cite{kkop} 
consists in chosing a linear field configuration, $\phi(y) = c_2|y|$ on the interval
$-y_{\rm m}<y<y_{\rm m}$ with end points identified. It was argued that in equlibrium
$(\partial_M \phi \partial^M\phi  - c_1^2)^2 = (b^{-2}(\phi')^2 - c_1^2)^2 =
(c_2^2/b^2-c_1^2)^2$ has to vanish, leading to a prefered value for the scale factor 
in the transverse direction, $b_0= c_2/c_1$, thus fixing the volume of 
the extra dimension, $L=2y_{\rm m}b_0$. The linear scalar field profile may 
originate from placing branes at $y=0$ and $y=y_{\rm m}$ and fixing the 
value of the scalar field on the branes. Unlike the Goldberger-Wise stabilization 
proposal \cite{GW}, the mechanism involving (\ref{intro}) does not lead to a backreaction 
on the background geometry. 
Incidentally, very similar constructions in the 4D setting \cite{ghost,therest} 
have been analyzed in detail recently as a dynamical way to break the Lorentz 
invariance in 4D.

%Such stabilization of an extra dimension 
%by "higgsing" the graviton in the transverse direction. 

This paper is organized as follows. Section 2 presents the solution to 
Einstein's and scalar field equations in 5D for both 
choices of sign for the kinetic term and investigates 
small deviations of the radion field from equilibrium. Section 4 
presents scalar field models that can form positive--negative tension brane 
pairs and the static solutions for the scale factor. 
Section 5 describes localization of gravity using scalar fields 
including a Gaussian profile solution that arises in the 
presence of the bulk tachyon. Section 6 contains discussion and conclusions.

\bigskip

\sect{Static solution in 5D in the presence of a bulk ghost}

In this section we consider a five-dimensional action in the presence
of a bulk scalar and two 3-branes located at points $y=y_1,y_2$ in the 5th dimension. 
\begin{eqnarray}\label{action}
S = - \int d^4x dy \sqrt{-g}\left\{ \fr{1}{2\kappa^2}\hat R +
 \fr{K(\phi)}{2}\partial_M \phi \partial^M \phi  + V_B(\phi) + V_1(\phi)\delta(y-y_1)
+V_2(\phi)\delta(y-y_2)\right\}.
\end{eqnarray}
$V_B(\phi)$ is the bulk potential, $V_{1(2)}$ are field-dependent brane 
tensions which can be either positive or negative and $K(\phi)$ is the kinetic 
function. If $K(\phi)$ is sign-definite and only one scalar field is present, 
then without loss of generality 
we can consider two choices,
\be
K=\left\{\begin{array}{c}+1 ~~~{\rm "normal"~scalar}\\-1 ~~~{\rm "ghost"~scalar}.
\end{array}\right.
\label{choice}
\ee

Throughout this paper we will consider only static solutions, and therefore
it suffices to choose the simplest ansatz for the metric and the scalar field:
\be 
ds^2 = a(y)^2 d x_4^2 + b^2 dy^2,~~~ 
\phi = \phi(y).
\ee

With the use of this ansatz, the five-dimensional field equations take a very simple form, 
\begin{eqnarray}
\fr{a''}{a} +\left(\fr {a'}{a}\right)^2 = \fr{\kappa^2}{3} T^i_i\\
2\left(\fr {a'}{a}\right)^2 = \fr{\kappa^2}{3}  T_5^5,
\end{eqnarray}
where index $i=1,2,3$ labels four-dimensional spatial components of tensors
and primes denote differentiation with respect to $y$-coordinate.  
The stress energy components can be expressed in terms of the parameters of 
the action (\ref{action}). Its bulk component is given by
\be
T_{MN} = K \partial_M \phi \partial_N \phi - g_{MN}
\left [ \fr K2\partial_P\phi\partial^P\phi + V_B(\phi)\right].
\ee
and at points $y_{1(2)}$ the brane contributions must be added.
Scalar field satisfies its equation of motion, 
\be
K\left(\phi'' +4\fr{a'}{a}\phi'\right) = \fr{\partial V(\phi)}{\partial \phi},\label{eom}
\ee 
where the potential includes bulk and brane contributions, 
$V(\phi)=V_B+\sum V_i\delta(y-y_i)$.

The solution for a "normal", $K=+1$, case is well known when the bulk potential
is a constant, $V_B(\phi) = \Lambda_B$. For the negative sign of $\Lambda_B$ 
the solution for the scale factor and the field profile in the bulk is given by \cite{kop1}
\be
\phi'(y) = {\rm const} \times a(y)^{-4};~~~a(y)^4= a_0^4 \sinh(\omega(y-y_0)) ~~~ {\rm for}~K=1,
\label{sinh}
\ee
where $\omega^2 = 8|\Lambda_B|\kappa^2/3$ and $b$ is set to 1.
Notice the monotonic behavior of $a(y)$ which means that the 
5-dimensional space ends either on the singularity at $y= y_0$, 
or on the negative tension brane. Solution (\ref{sinh}) has to be supplemented 
by the boundary conditions for $\phi$, $\phi'$ and $a'$ at $y=y_1,y_2$ 
which are listed in Refs. \cite{kop1,kop2}.

It is important to emphasize that in the absence of the negative tension branes 
the absence of a minimum in the scale factor is, in fact, a general 
consequence of Einstein's equations and the form of the stress energy tensor with $K=+1$.
This property does not depend on the particular choice of bulk potential $V_B(\phi)$
as any possible choice of $V(\phi)$ would not be able to 
create a minimum in the scale factor. Indeed, suppose that the scale factor has an extremum 
at $y=y_0$. Near the extremum, $a = a_0$, $a' = 0$, and 
$\phi(y_0)=\phi_0$. The l.h.s. of the 55 Einstein's equation goes to zero near 
$y=y_0$, and so is $T^5_5$. For the standard sign of 
the kinetic term, $K=+1$, $T_5^5 = \phi'^2/2 - V_B(\phi_0)=0$ which immediately gives 
$\phi'^2 = 2V_B(\phi_0)$. Therefore, near the extremum $V_B(\phi_0)>0$ and $T^i_i=
-\phi'^2/2 -V_B(\phi_0)<0$. Using the $ii$ equation, we immediately conclude that $a''<0$, 
which may correspond only to the {\em maximum} of the scale factor. This 
conclusion generalizes to an arbitrary number of the scalar fields, and therefore 
the minimum of the scale factor cannot be achieved for {\em any} choice of potential(s)
and bulk matter content if the sign of all 
kinetic terms is standard. In this proof, we used the continuity
of $a'$ and $\phi'$ at $y_0$, and the inclusion of the negative tension brane 
allows to escape these no-go arguments by arranging a discontinuity in $a'$.

On the other hand, the choice of $K = -1$ would naturally lead to the 
{\em minimum} of the scale factor.  Indeed, it is easy to show that the similar 
chain of arguments leads to the conclusion that for the bulk ghost the extremum 
point in  the scale factor corresponds to $a''>0$. 

It is a natural point of view to consider infinitely thin branes as generalizations
of classical domain wall solutions in the limit of small thickness. It is 
possible to construct the generalization of the positive tension 
brane as a domain wall solution for the scalar field with $K=1$ \cite{Gregory,KKS}. 
At the same time, any attempt 
to "resolve" a negative tension brane as the classical solution of fields with $K=1$ will 
necessarily fail because of the arguments presented above. One can try 
to resolve the negative tension brane by giving it a finite width without assuming 
a definite sign of the kinetic term for the scalar field. Then, immediately, one would 
be forced to conclude that in the middle of such brane the contribution 
of the ghost field must be dominant, and therefore it is reasonable to expect that 
the negative tension brane must be "constructed" from the ghost field. 

%%%%%%%%%%%%%%%%%%%%%%%
\begin{figure}
 \centerline{%
   \includegraphics[width=12cm]{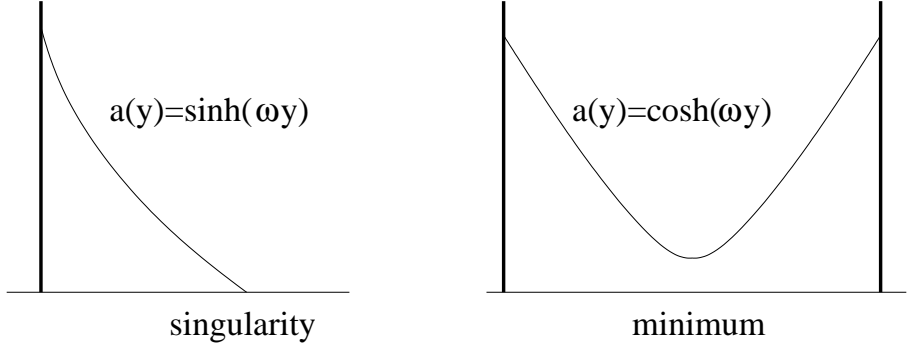}%
         }
\vspace{0.1in}
 \caption{Dependence of the scale factor on the sign of the kinetic term of 
the bulk scalar: $a(y)=\sinh(\omega y)$ for $K=1$ and  $a(y)=\cosh(\omega y)$ for $K=-1$.}
\end{figure}
%%%%%%%%%%%%%%%%%%%%%%%

Returning to the case of the trivial bulk potential, $V_B(\phi)= \Lambda_B$, 
we find easily the exact static solution for the metric and the bulk 
field,
\be
\phi'(y) = \pm\sqrt{|\Lambda_B|}a(y)^{-4};
~~~a(y)^4= a_0^4 \cosh(\omega y) ~~~ {\rm for}~K=-1,
\label{cosh}
\ee
with the same choice of $\omega$ as before. This solution does not have any 
singularity in $\phi(y)$ or $a(y)$. This way of avoiding a singularity 
is similar in spirit to the introduction of the higher-derivative terms in the action \cite{Bob}, 
but simpler in relaization. 
As argued above, the existence of ghost leads to the upturn of the 
scale factor and therefore, in some sense, the ghost field stress energy 
plays a role of the effective negative tension brane. 

Placing branes on the both sides from the minimum, gives a static model 
of the fifth dimension with two positive tension branes, which is impossible 
for the standard choice of the kinetic term in the bulk. The boundary conditions besides
a continuity condition for $\phi$ include four conditions 
for the scale factor and $\phi'$ at the positions of the branes,
\be
\fr{[a']}{a_i}\Big|_{y=y_i} = (-1)^i\fr{\kappa^2}{3}V_i(\phi)\Big|_{y=y_i}, ~~~ 
[\phi']\Big|_{y=y_i} = (-1)^i\fr{\partial V_i(\phi)}{\partial \phi}\Big|_{y=y_i}, 
\label{boundary}
\ee
where $i=1,2$ and $[..]$ denotes a jump across the brane. 
This is an over-determined set of equations that requires additional
tuning of the parameters in order to produce a static 4D-flat metric. 

At this point we would like to study the stability of the resulting two-brane configuration
with respect to small fluctuations in the size of extra dimension. Following the 
method of Ref. \cite{kop2}, we calculate the second variation of the effective action 
over the scale factor $b$ and use boundary conditions to simplify the answer. 
The result can be presented as a sum of two terms evaluated at 
the position of the branes, 
\be
b^2\fr{d V_{\rm eff}}{db^2} = \fr{\kappa^2}{3} \sum_{i=1,2} (y_ib)^2 a_i^4 V_i 
\left( \fr {\partial V_i}{\partial \phi}\right)^2.
\label{radionmass}
\ee
Interestingly enough, this expression is always positive, and therefore the variation in
the size of extra dimension does not lead to an instability. This situation 
is different from the $\sinh$-type solution where both stable and unstable 
configurations are possible \cite{kop2}. 

The positivity of the radion mass squared does not guarantee the overall stability of
this construction because the system with ghosts can always lead to an instability, 
as the energy is not bounded from below. In other words, the 5D ghost will 
produce a 4D ghost in an effective theory. A possible remedy to this 
problem can come either through higher dimensional 
operators in the bulk  $ (\partial_M\phi \partial^M\phi)^2$ and alike with $K=1$ 
or sufficiently large "induced" kinetic terms on the brane that in  the effective 4D theory can 
overpower the ghost-like sign from the bulk,
\be
S_{\rm ind} =-\sum_i\int \sqrt{-g}d^4xdy \delta(y-y_i)K_i\partial_\mu\phi\partial^\mu \phi .
\ee
where both $K_i>0$ and $\mu$ runs over 4D indices.

\sect{Modeling branes and extra dimension}

In this section we consider the bulk scalar field(s) with alternating sign of the
kinetic term. If $K=K(\phi)$ changes the sign, the positive(negative) sign domain can play a role 
of the positive(negative) tension branes. As a result, we can construct a 
continuous field configuration in the compact fifth dimension that can be viewed 
as a positive tension brane -- negative tension brane system. 

To demonstrate such possibility we go backwards, and choose the following 
scale factor and $\phi$ configurations,  
\be
a = a_0 \exp\{A_0 \cos(\omega y)\};~~~ 
\phi(y) = \phi_0 \cos(\omega y).
\label{config}
\ee
We consider this solution on the interval $y= 0$ and $y= 2\pi/\omega$  with the end 
points identified. If $A_0\gg 1$ it is analogous to a smooth version of the Randall-Sundrum metric .

Plugging (\ref{config})to Einstein's equations, we reconstruct 
the scalar field Lagrangian consistent with this solution,
\be
{\cal L}_{\rm scalar} = -\fr{M_*^3}{2}(\partial_M \chi)^2\fr{\chi}{1-\chi^2} +
|\Lambda_B|\left( 1- \fr{3}{4M_*^2\kappa^2}\chi - \chi^2\right).
\label{strange}
\ee
In this Lagrangian, we introduced a rescaled field $\chi$, 
$\chi = \phi/(\phi_0M_*)$. As we expected, the kinetic term 
is not sign-definite. Near the maximum(minimum) of the scale 
factor the kinetic term in (\ref{strange}) 
goes to $-1(1)$. 

The warp factor is given by $a_{\rm max}/a_{\rm min} = \exp\{2A_0\}$, where
$A_0$ is determined via the relation $A_0 = M_*^3\kappa^2/3$, and $\omega$ is given by
$\omega = 3|\Lambda_B|/(2\kappa^2M_*^6)$. It is large 
if $M_*$ is larger than the 5D Planck mass $M_{(5)}$. The "effective"
branes ar located at $y = 0$ and $y = \pi/\omega$. Their width $L$
can be estimated from the condition $a_{\rm max}/a(L)\sim 1$ which is equivalent
to $\omega L \sim A_0^{-1/2}$, with the same estimate for the minimum of the 
scale factor. Therefore, in the limit of large warping, the 
effective brane widths are small. 

Two bulk scalar fields with opposite signs of the kinetic terms allow for the
similar solution  with a much simpler scalar Lagrangian. Assuming the existence of 
a scalar field $\phi$ with $K=1$ and ghost field $\chi$ with $K=-1$, we 
observe that the scalar Lagrangian 
\begin{eqnarray}\nonumber
{\cal L}_{\rm scalar} = -\fr 12 (\partial_M \phi)^2 +\fr 12(\partial_M \chi)^2 -
\lambda_4(\phi^4+ \chi^4) - \alpha \phi^2\chi^2-\lambda_2\phi^2-\mu_2\chi^2 -\Lambda_B,
\label{phichi} 
\end{eqnarray}
is consistent with the following solution, 
\be
a = a_0 \exp\{A_0 \cos(2\omega y)\}; ~~~ \phi(y) = \phi_0\cos(\omega y);~~~
\chi(y) = \phi_0\sin(\omega y),
\label{sincos}
\ee
under the following set of conditions:
\begin{eqnarray}
\omega^2 &=& \mu_2-\lambda_2=-\fr{1}{4}(\mu_2+\lambda_2 +2\alpha), ~~~(\mu_2>\lambda_2)
\nonumber\\\label{conditions}
\fr{\kappa^2}{3}\omega^2&=&\lambda_4-\fr \alpha2\\\nonumber
\fr{12}{\kappa^2}A_0 &=& -\phi_0^2 = \fr{1}{2\lambda_4}\left(\fr{\lambda_2+\mu_2}{2}+
\sqrt{\left(\fr{\lambda_2+\mu_2}{2}\right)^2 -4\Lambda_b\lambda_4}\right).
\end{eqnarray}
These conditions necessitate fine tuning between the parameters of Lagrangian (\ref{phichi}).
As in the previous example, the dimension can be considered compact with the 
negative tension branes built from $\chi$ and positive tension branes from $\phi$.

The analysis of the stability of the resulting solutions goes beyond the scope 
of the present paper. We note, however, that the solutions obtained with the use of a 
sign-indefinite $K(\chi)$ similar to Eq. (\ref{strange}) can avoid a ghost 
in the effective 4D theory. This is because the integral of $a^2K(\chi)$ 
over the compact dimension determining the sign of an {\em effective} $K$ 
for the 4D scalar may turn out to be positive. 

\sect{Localizing gravity using ghosts and tachyons. Gaussian extra dimension.}

Gravity can be localized in four dimensions even if the fifth 
dimension is not compact \cite{RS,KR}. In this section we 
search for a scalar field configurations including ghosts and tachyons  
that are capable of localizing gravity. 

All examples in the previous sections operated with the compact dimension,
which leads to an over-determined set of the boundary conditions and as a 
consequence the fine tuning among the parameters of the 5D Lagrangian that ensures 
the absence of the 4D cosmological constant. This situation may change if 
the dimension is not compact and the boundary conditions for the scalar 
field at $y=\pm\infty$ are not imposed. 

The absence of fine tuning can be exemplified in one-brane modification 
of the solution (\ref{cosh}). If a single positive tension brane is located at
$y=0$, the solution for the scale factor to the right of the brane would take 
the form $a^4 = a_0^4 \cosh(\omega(y-y_{\rm min}))$ with its symmetric reflection 
to the left of the brane. For simplicity, let us assume the linear dependence 
of the brane tension on the value of the scalar field, $V_{\rm br}(\phi) = 
\Lambda_0 +\lambda\phi$. The boundary condition for the derivative of the 
scalar field  and of the scale factor on the brane, 
\begin{eqnarray}
\label{boundaryalpha}
\phi'(0) = \fr{|\Lambda_B|^{1/2}}{2\cosh(\omega y_{\rm min})}= \lambda,\\
4\fr{a'(0)}{a_0} = \tanh(\omega y_{\rm min}) = \fr{2\kappa^2}{3}
\left(\Lambda_0 + \lambda \phi_0\right),
\nonumber
\end{eqnarray}
can be easily satisfied for {\em any} values of the parameters $\Lambda_B$, $\Lambda_0$ 
and $\lambda$. This is because the boundary conditions for $\phi$  are not imposed at $y=\pm \infty$.
This example uses the ghost field and  is different 
from the known case with the regular ($K=1$) scalar field in which
the solution for $\phi(y)$ and $a(y)$ end on the singularity and where the 
boundary conditions cannot be ignored \cite{Nilles}. If the coupling constant of the 
ghost to the brane is small, then a large warping between $y=0$ and $y=y_{\rm min}$ 
can be achieved, $\cosh(\omega y_{\rm min})= |\Lambda_B|^{1/2}/(2\lambda)$. In this case, 
it is reasonable to expect that the 4D graviton can be quasi-localized, with the 4D behavior
within certain radius.   

To consider a complete localization of gravity on an RS-type configuration, 
we would like generalize the scalar Lagrangian to include a nonlinear kinetic term,  
\be
{\cal L}_{\rm bulk} = - K (\partial_M \phi \partial^M \phi)^n - \Lambda_B
\label{generalisation}
\ee
where $n$ is an arbitrary, not necessarily positive or integer number. As before, 
$K$ can take both signs.

Exact static solutions can be found for negative $\Lambda_B$, for both signs of $K$, 
and any $n$, 
\ba
\nonumber
a(y) = a_0 
\left [ \cosh (\omega y) \right ]^{\fr 12 -\fr{1}{4n}}
~~~{\rm for}~K=1, n<\fr 12~{\rm and}~K=-1, n>1/2\\
a(y) = a_0 
\left [ \sinh (\omega y) \right ]^{\fr 12 -\fr{1}{4n}}
~~~{\rm for}~K=1, n>\fr 12~{\rm and}~K=-1, n<1/2\\
\phi' = {\rm const}\times a^{-4/(2n-1)},
\nonumber
\ea
with the same relation between $\omega$ and $\Lambda_B$ as before. 
Obviously, the solution with the $\sinh$ function is defined only where it is positive. 

Choosing $K=1$ and $0<n<1/2$ leads to a complete localization of gravity, 
because $a(y) = a_0/\cosh^\beta(\omega y)$ with $\beta = \fr 12 -\fr{1}{4n}>0$
has an asymptotic behavior of $a(y) \sim \exp\{-\beta\omega|y|\}$ at $\beta\omega|y|\gg1$. 
Consequently, the 4D Planck mass scales as $M^{-2}_{(4)} \sim \kappa^2\beta\omega$ 
and can be calculated exactly if needed. This model may or may not have a positive tension 
brane included. Indeed, a solution without a brane is perfectly sensible with the scalar
field and the scale factor continuous at all $y$. Notice that the effective 4D
theory obtained from (\ref{generalisation}) with $K=1$ and $0<n<1/2$ contains 
a "normal" 4D scalar with $K=1$. This solution can therefore be 
considered under a category of "self-tuning" solutions,
as for any  $\Lambda_B$ and $\Lambda_{0}$ and $\alpha$ 
there exists a static solution without the need to fine-tune 
the parameters of the 5D Lagrangain. Again, this is because
we do not impose any boundary conditions for $\phi$ at infinity. 
By choosing $n= 1/3$ the dependence of the scale factor on $y$ can be made exactly the same as in 
the models \cite{KKL} where a more complicated construction with the use of a 
non-standard Lagrangian for the four-form field was suggested. The correspondence 
between the scalar case with $n=1/3$ and the four-form field construction was found earlier in 
Ref. \cite{dual}. We would refrain 
from speculations whether such solutions lead to a solution to the cosmological 
constant problem, because it is not clear whether a small perturbation around 
a static solution would tend to converge to a flat solution with time as it is 
generally not the case \cite{Medved}. 

Finally, we consider a possibility for a localization of the 4D graviton using the 
{\em Gaussian} $y$-profile for the scale factor. In some sense, this localization 
is stronger than in RS-type models where asymptotically $a(y) \sim \exp\{-\beta\omega|y|\}$. 
It turns out that a tachyon field configuration is capable of 
creating a $a = a_0 \exp\{- \fr 12 \alpha y^2\}$ profile.

Indeed, let us consider a scalar Lagrangian with $K=1$ and a tachyonic mass term,
$m^2 = - m_t^2<0$, and a positive cosmological constant $\Lambda_B$. 
\be
{\cal L}_{\rm scalar} = -\fr 12 \partial_M \phi \partial^M\phi 
- \Lambda_B + \fr 12 m^2_t\phi^2.
\label{tachyon}
\ee
Using this Lagrangian we find the following simple solution,
\be
\phi(y) = c y, ~~~a(y) = a_0\exp\{- \fr 12 \alpha y^2\},
\label{gauss}
\ee
where $\alpha$ and $c$ can be found from the following relations,
\be
c =\pm 2 \sqrt \Lambda_B, ~~~ \alpha = m^2_t/4,
\ee
subject to an additional fine-tuning between the parameters of (\ref{tachyon}):
\be
m^2_t = \fr{8\kappa^2}{3} \Lambda_B.
\ee

Using these relations, we find an effective (inverse) 4D Newton constant,
\be
\fr{1}{\kappa_{(4)}^2}= \fr {1}{\kappa^2}\int \fr{a^2}{a_0^2}dy 
= \fr{2\sqrt\pi}{\kappa^2 m_t}.
\label{Newton}
\ee
The gravity is essentially localized in the vicinity of $|y|<m_t^{-1}$. The 
Kaluza-Klein spectrum is continuous, and the wave functions of KK gravitons 
are suppressed near $y=0$. The exact study of the excited modes can be done, 
following the RS analysis, but this goes beyond the scope of the present paper. 

The point of $\phi = 0$ may happen at arbitrary $y$, as the whole 
solution (\ref{gauss}) can be translated by an arbitrary shift,
$y\rightarrow y+d$. Such symmetry may be broken if there is an additional 
discrete symmetry $\phi(-y)=-\phi(y)$. This "orbifolding" symmetry may 
be a welcome addition to (\ref{tachyon}) as it projects out the 
tachyonic mode from  an effective 4D theory. Solution (\ref{gauss}) 
may be considered as a limiting case of a more generic situation where 
$\lambda_4\phi^4$ term is present in the potential. In this, solution 
(\ref{gauss}) would describe an interior of the domain wall, with the 
scalar field being approximately linear with $y$ and where the quartic term is small. 
Trapping of gravity on field configurations with quartic interaction has been discussed in a recent 
paper \cite{deCarlos}.

\sect{Discussion and Conclusion}

In this paper we showed that the bulk ghosts generically lead to the {\em minimum} 
in the scale factor. We use this observation to argue that ghost fields can be 
thought of as "substitutes" for the negative tension branes. A bulk 
ghost and a negative bulk cosmological constant $\Lambda_B$ 
create the $\cosh$-like profile of the scale factor in the bulk, $a(y)^4\sim \cosh(\omega y)$.
Therefore, in this model the 5th dimension can be made compact and
4D flat by introduction of only positive tension branes. We analyzed the stability of
the resulting $\cosh$-like configuration with respect to small fluctuations of the transverse
scale factor $b$ (or radion) and found that the model is stable for any position 
of the branes and any tensions. Of course, the issue of the stability for the whole construction
depends on the fate of ghost in the 4D effective theory. 
Generically, a 5D ghost would produce a ghost in 4D, which then would be 
unacceptable. A possible cure can come from the higher dimensional 
operators and/or from the induced kinetic terms for $\phi$ on the 
branes. 

Another possibility of {\em not} having a ghost in the effective 4D theory 
is the case of a scalar field with variable sign of the kinetic term in the 
bulk. As we have shown, such field can be used to model extra dimension and branes. 
This way we obtained a smooth generalization of the RS metric, $a = a_0 \exp\{A_0
\sin(\omega y)\}$, with "effective" branes of $\sim A_0^{-1/2}$ widths located at the
minimum and the maximum of the scale factor. 

Models with compact dimension always contain the overdetermined 
set of boundary conditions leading to the fine-tuning among 
the parameters of the scalar Lagrangian needed to maintain 
an (approximate) 4D flatness. In some models with non-compact dimension
the over-determined set of conditions can be avoided by {\em not} 
imposing them on the scalar field at infinity. Introducing a 
non-linear (and non-analytic) kinetic term for the scalar 
field, we find that the solutions of $a\sim 1/\cosh^\beta(\omega y)$ become 
possible, whereas scalar field is free to go to infinity at infinite $y$. 
Such models lead to the trapping of gravity on these scalar field configurations,
and can be achieved without additional fine-tuning, and do not necessitate the
presence of brane(s). 

We have also shown that the 5D model with the tachyonic mass term 
can lead to a stronger localization of gravity by creating a Gaussian 
profile for the scale factor in the 5th dimension and a linear profile
for the scalar field. 4D flatness of the solution requires 
one fine-tuning between the mass of the tachyon and the (positive)
bulk cosomological constant. A 4D tachyon can be projected
out by introducing an additional symmetry, $\phi(-y)\rightarrow -\phi(y)$. 
However, at this point it is premature to say whether the 
trapping of gravity on a tachyonic field configuration can be used in 
a more content-rich model, such as {\em i.e.} bosonic string theory.

%%%%%%%%%%%%%%%%%%%%%%%%%%%%%%%%%%%%%%%%%%%%%%%%%%%%%%%%%%%%%%%%%%%%%%%%%%%%%%%%

{\bf Acknowledgements} 

This work has been supported in  part by the NSERC of Canada.

\end{document}